\documentclass[12pt]{iopart}
\usepackage{graphicx}
\usepackage[utf8x]{inputenc}
\usepackage[usenames,dvipsnames]{xcolor}
\usepackage{soul}
\usepackage{enumitem}

\begin{document}


\title{Identity of the JET M-mode and the ASDEX Upgrade I-phase phenomena}

\author{D. I. R\'efy$^{1}$, E. R. Solano$^{2}$, N. Vianello$^{3}$, S. Zoletnik$^{1}$, D. Dunai$^{1}$, B. Tál$^{1}$, M. Brix$^{4}$, R. Gomes$^{5}$, G. Birkenmeier$^{6}$, E. Wolfrum$^{6}$, F. Laggner$^{7}$, M. Griener$^{6}$, O. Asztalos$^{8}$, E. Delabie$^{9}$, ASDEX Upgrade Team, JET Contributors\footnote{See the author list of  “Overview of the JET preparation for Deuterium-Tritium Operation” by E. Joffrin et al. to be published in Nuclear Fusion Special issue: overview and summary reports from the 27th Fusion Energy Conference (Ahmedabad, India, 22-27 October 2018)}, EUROfusion MST1\footnote{See author list of “B. Labit et al 2019 Nucl. Fusion 59 086020 (https://doi.org/10.1088/1741-4326/ab2211)''}}

\address{$^1$Wigner RCP, Budapest, Hungary}
\address{$^2$Laboratorio Nacional de Fusi\'on, CIEMAT, Madrid, Spain}
\address{$^3$Consorzio RFX Padova, Italy}
\address{$^4$CCFE, Abingdon, United Kingdom}
\address{$^5$IPFN, Instituto Superior Técnico, Universidade de Lisboa, Lisboa, Portugal}
\address{$^6$MPI for Plasma Physics, Garching, Germany}
\address{$^7$Princeton Plasma Physics Laboratory, Princeton, New Jersey 08543, USA}
\address{$^8$Budapest University of Technology, Budapest, Hungary}
\address{$^9$ORNL, Tennessee, USA}


\ead{refy.daniel@wigner.mta.hu}


\begin{abstract}
An H-mode plasma state free of edge-localized mode (ELM), close to the L-H transition with clear density and temperature pedestal has been observed both at the Joint European Torus (JET) and at the ASDEX Upgrade (AUG) tokamaks usually identified by a low frequency (LFO, 1-2 kHz), m=1, n=0 oscillation of the magnetics and the modulation of pedestal profiles. The regime at JET is referred to as M-mode while at AUG as intermediate phase or I-phase. This contribution aims at a comparative analysis of these phenomena in terms of the density and temperature pedestal properties, the magnetic oscillations and symmetries. Lithium beam emission spectroscopy (Li-BES) and reflectometer measurements at JET and AUG show that the M-mode and the I-phase modulates the plasma edge density. 
A high frequency oscillation of the magnetics and the density at the pedestal is also associated with both the M-mode and the I-phase, and its power is modulated with the LFO frequency. The power modulation happens simultaneously in every Mirnov coil signal where it can be detected.
The bursts of the magnetic signals and the density at the pedestal region are followed by the flattening of the density profile, and by a radially outward propagating density pulse in the scrape-off layer (SOL). The analysis of the helium line ratio spectroscopy (He-BES) signals at AUG revealed that the electron temperature is modulated in phase with the density, thus the I-phase modulates the pressure profile gradient. This analysis gave opportunity to compare Li-BES and He-BES density profiles at different locations suggesting a toroidal and poloidal symmetry of the density modulation. The presented results indicate that the regimes, the AUG I-phase and the JET M-mode, exhibit similar characteristics, which leads to the conclusion that the regimes are likely the same.
\end{abstract}



  \submitto{\NF}



\section{Introduction}
\label{sec:intro}

A high confinement mode (H-mode) plasma state free of Type-I edge-localized modes (ELMs), close to the L-H transition with clear density pedestal has been observed both at the Joint European Torus (JET) and at the ASDEX Upgrade (AUG) tokamaks usually identified by a low frequency (1-2 kHz), m=1, n=0 oscillation (LFO) of the magnetics and the modulation of pedestal profiles. The observed axisymmetric magnetic oscillation has an odd parity across the mid-plane: it is magnetically similar to an up-down oscillation of the plasma. The plasma state is referred to as M-mode~\cite{Solano2017} at JET while as intermediate phase (I-phase~\cite{Conway2011}\cite{Birkenmeier2016b}\cite{Manz2016}\cite{Medvedeva2017}\cite{Cavedon2017}) at AUG. 

Similar phenomena which exhibit oscillations between different phases, often called Limit Cycle Oscillations (LCOs), have been reported in various devices: COMPASS~\cite{Grover2018}, DIII-D~\cite{Colchin2002}\cite{Schmitz2014}, EAST~\cite{Xu2011}\cite{Xu2014}\cite{Xu2014b}, HL-2A~\cite{Xu2015}, JFT-2M~\cite{Kobayashi2014a} and TJ-II~\cite{Estrada2015}. These appear close to the L-H transition and are considered as predator-prey dynamics in the flow - turbulence - profile system~\cite{Schmitz2017}\cite{Cheng2013}. 

The I-phase is probably not governed by predator-prey dynamics as was shown in reference~\cite{Birkenmeier2016b} and \cite{Cavedon2017}, however the closed cycles in the Lissajous plots between $\mathbf{E_{r}}$ and the density fluctuations during the early I-phase, shown in reference~\cite{Medvedeva2017} indicates that this option cannot be excluded. The growing high frequency precursor in $\mathbf{\dot{B}_{r}}$ at AUG in reference~\cite{Birkenmeier2016b} shows clear electromagnetic nature of the I-phase. The n=0, m=1 LFO of the magnetics can be the consequence of the currents in the SOL, which slightly modify the equilibrium, however, the poloidal Alfvén wave frequency scaling as shown in reference~\cite{Solano2017} along with the stationary frequency of the mode suggests that an MHD wave is excited during the M-mode. The I-phase precursors are similar to Type-III ELM precursors, and the I-phase bursts are considered pre-mature ELMs in reference~\cite{Birkenmeier2016b}, however the I-phase frequency does not scale with heating power, while Type-III ELM frequency scales inversely with heating power. An alternative explanation for the $n_{e}$ low frequency oscillation was shown in reference~\cite{Clairet2018} claiming that such modulation can be the consequence of tilted turbulent structures moving due to the $\mathbf{E \times B}$ flow, based on fast frequency swept X-mode reflectometer measurements both at JET and AUG. While the interpretation of the results is different from that we propose, namely that the flattening of the density profile can cause such reflectometer spectra, it has to be emphasized that the same pedestal dynamics is observed by an independent diagnostic at both machines. Beside these open questions there is a broad consensus that the M-mode and the I-phase are part of the H-mode due to their improved confinement properties and formation of the temperature and density pedestal.


While present fusion experiments can be operated considerably above the L-H transition power threshold, ITER probably will be operated with marginal heating power~\cite{Martin2008}. The M-mode and the I-phase are present in such circumstances, and at least the AUG I-phase is present in every discharge with favorable ion $\mathbf{\nabla B}$ drift direction ($\mathbf{B \times \nabla B}$ pointing towards the X-point) and with an L-H transition. Considering that large ELMs must be avoided at ITER to preserve integrity of plasma facing components, an H-mode plasma regime without ELMs and with reasonable confinement would be an appropriate plasma operation scenario for ITER. No special techniques, like reversed magnetic field, high plasma rotation shear nor magnetic perturbation coils are needed to enter the M-mode or the I-phase which will be either impossible or very hard to make at ITER. Another important feature is that they emerge independently of the heating technique.

The appearance of such oscillations at the L-H transition at various machines with different heating methods and plasma parameters indicates that this might be a common, general phenomenon. A comparative analysis of the M-mode and the I-phase in terms of low and high frequency magnetic oscillations, high frequency oscillation power modulation, density and temperature pedestal properties and high frequency density oscillations is presented in this paper.

The paper is organized as follows. Section~\ref{Sect.diag} provides a brief description of the lithium beam emission spectroscopy (Li-BES, section~\ref{Sect.Li-BES}) systems at JET and AUG, and of the helium line ratio spectroscopy diagnostic (He-BES, section~\ref{Sect.He-BES}) at AUG. These diagnostics were used for the main part of the presented analysis beside of the Mirnov coils. Section~\ref{Sect.dataset} focuses on the properties of the analyzed discharges. The high frequency oscillations (HFO, section~\ref{Sect.HFO}) as well as the high frequency oscillation power modulation (HFO power modulation, section~\ref{Sect.HFB}) of the magnetic signals as the most general property of the phenomena, along with the toroidal symmetry properties of the LFO and the HFO (section~\ref{Sect.modenumber}) are discussed in section~\ref{Sect.magnetic}. The results of the $n_{e}$ profile oscillations studies are presented in section~\ref{Sect.densLFO}, combined $n_{e}$ and the $T_{e}$ profile characterization is shown for AUG in section~\ref{Sect.presLFO}, while the JET pedestal temperature fluctuations are investigated in section~\ref{Sect.tempLFO}. The high frequency density modulation properties of the modes are shown in section~\ref{Sect.denshfb}, while the insights of the Li-BES fluctuation measurement interpretation and modeling can be found in \ref{App.fluct1} and \ref{App.fluct2}. The results are summarized in section~\ref{Sect.summary}, indicating that the physics background of the M-mode and the I-phase are the same.

\section{Utilized diagnostics \label{Sect.diag}}
\subsection{Lithium Beam Emission Spectroscopy \label{Sect.Li-BES}}

Li-BES and reflectometer measurements at JET~\cite{Solano2017} and AUG~\cite{Birkenmeier2016b} show that the M-mode and the I-phase modulate the plasma edge density. The investigation of the density profile dynamics during these phenomena became possible with the upgraded Li-BES at both machines as the diagnostics are capable of density profile measurements up to the pedestal top with 0.5 - 1 cm spatial and 50 - 100 $\mu$s temporal resolution. The Li-BES technique is a routinely used diagnostic for electron density profile measurement at several plasma experiments~\cite{Zoletnik2018}, and works as follows. An accelerated atomic beam is injected into the plasma, where the beam atoms are excited and ionized by plasma particles. The ionization process results in a gradual loss of the atoms in the beam. The beam attenuation is such that the beam can penetrate only the edge of the plasma, thus Li-BES systems are used for electron density profile and fluctuation measurement of the outer plasma regions only, namely the plasma edge and scrape off layer (SOL). Spontaneous de-excitation of the beam atoms results in a characteristic photon emission which may be detected through an optical system. The distribution of the light emission along the beam (light profile) is measured by a detector system, from which the electron density distribution (density profile) is calculated~\cite{Schweinzer1992}\cite{Fischer2008}.

While the same Li-BES technique was utilized at both machines, the two diagnostics have slightly different properties. 
\begin{itemize}
\item
The JET Li-BES system~\cite{Refy2018} consists of a beam injector aligned vertically at the top of the tokamak, shooting downwards and a beam emission detection system. The observation range of the detection system can be set by a turnable mirror in the limiter shadow displaced toroidally from the beam, looking quasi parallel with the field lines. The beam emission is imaged on an array of 1 mm diameter optical fibers through a lens resulting in 6-10 mm spot size at the beam location, which defines the optical resolution of the system. The light from the fiber array is imaged on a 32 channel avalanche photo diode camera (APDCAM~\cite{Dunai2010}) in an optical enclosure located in the diagnostic hall, sampling the filtered light emission with 500 kHz. The beam can be modulated up to 10 kHz, thus the background radiation can be subtracted on the 50 $\mu$s timescale. The relative calibration of the channels is done by matching the beam emission signal in the plasma measured simultaneously by a spectrometer. The spectrometer measures the same input fibers as the APDCAM, utilizes spectral background subtraction and residual gas shot calibration. The density profile is calculated by a Bayesian algorithm~\cite{Fischer2008} from the background corrected and relatively calibrated beam emission signal. 
\item
The AUG system~\cite{Willensdorfer2014} consists of a beam injector aligned horizontally, shooting 30 cm above the mid-plane of the tokamak, and a beam emission detection system. The observation range is fixed, and defined by an array of lenses in an optical head displaced toroidally from the beam, looking quasi parallel with the field lines. Each elliptical lens images 6 $\times$ 12 mm of the beam on a circular fiber, which defines the optical resolution of the system. The light from each fiber is measured by a photo-multiplier after optical filtering around the Doppler shifted 670.8 nm Lithium line. The signal of the photo-multiplier is sampled with 200 kHz. The beam extraction can be modulated up to 250 Hz, and the background is subtracted on the 4 ms time scale accordingly, however, the modulation is operated most of the time the way, that the beam is on for 56 ms and off for 24 ms. The relative calibration of the channels is done by shooting into neutral gas where the beam emission is considered homogeneous along the beam, taking some beam attenuation into account. The density profile is calculated by a Bayesian algorithm~\cite{Fischer2008} from the light profile.
\end{itemize}

\subsection{Helium line ratio spectroscopy \label{Sect.He-BES}}

A new helium line ratio spectroscopy diagnostic (He-BES for brevity) is available~\cite{Griener2018} at the AUG tokamak, measuring the SOL and the plasma edge electron density and electron temperature profile simultaneously with up to 900 kHz temporal and up to 3 mm spatial resolution. A thermal helium beam is injected about 15 cm below the mid-plane of the AUG tokamak, and the line emission of the helium atoms is imaged on a fiber array which is then measured by a polychromator system at 4 wavelengths. The background emission is monitored with the same system by chopping the gas injection. The temperature and the density value can be calculated from the singlet/triplet (s/t) and the singlet/singlet (s/s) line ratio map, since the s/t is mostly temperature, while the s/s is mostly density dependent~\cite{Griener2018a}.

\section{Analyzed discharges \label{Sect.dataset}}

\begin{figure}
\includegraphics[width=\linewidth]{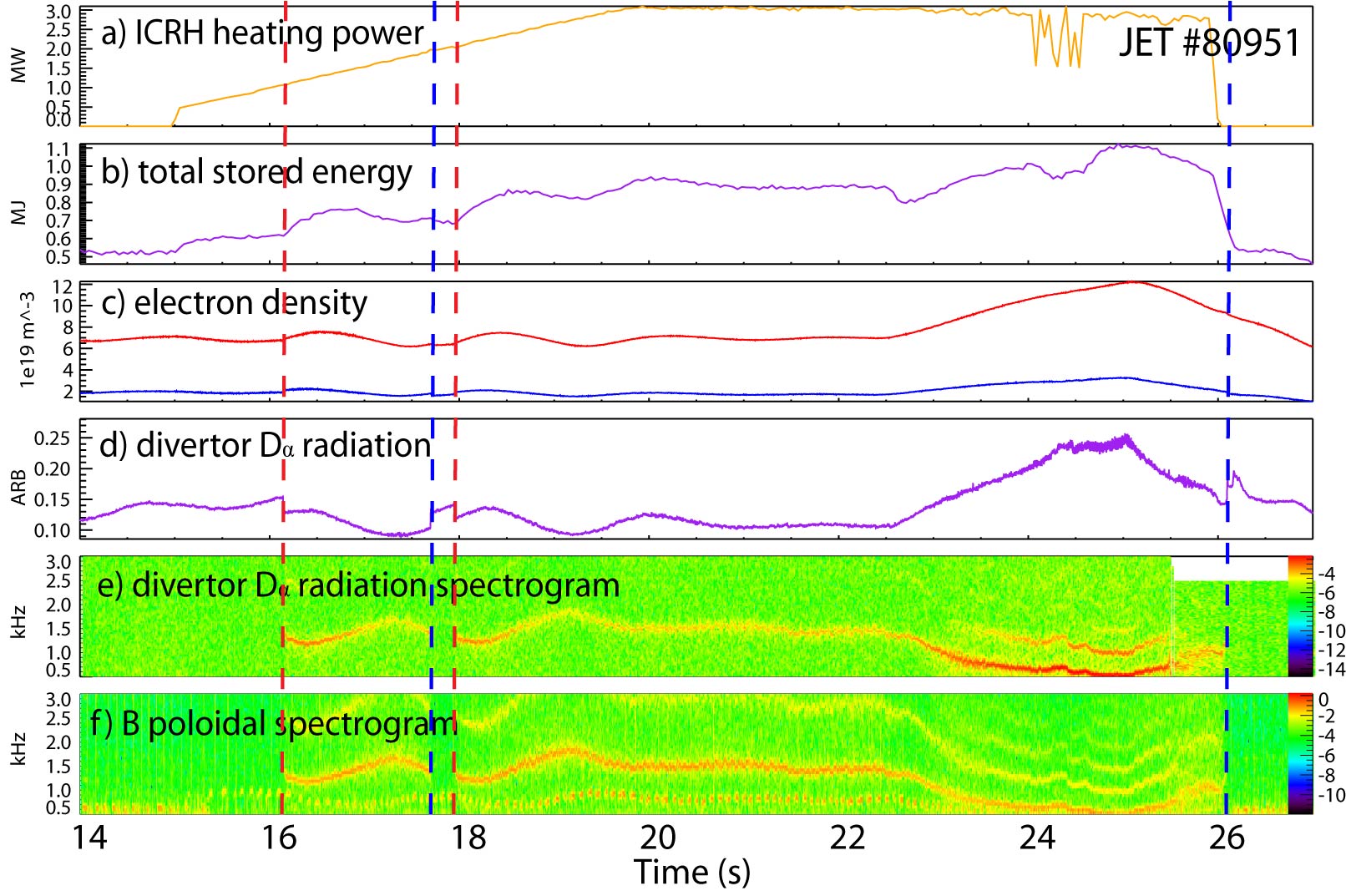} 
\caption{Overview of the analyzed JET discharge \#80951: a) ICRH heating power, b) total energy of the plasma, c) line integrated electron density from interferometer, blue - edge, red - core, d) divertor D$_\alpha$ radiation, e) spectrogram of the divertor D$_\alpha$ radiation signal, f) spectrogram of the poloidal magnetic field signal. The L-M/M-L transition times are indicated with a vertical red/blue dashed lines. \label{fig:jet.over2}}
\end{figure}

\begin{figure}
\includegraphics[width=\linewidth]{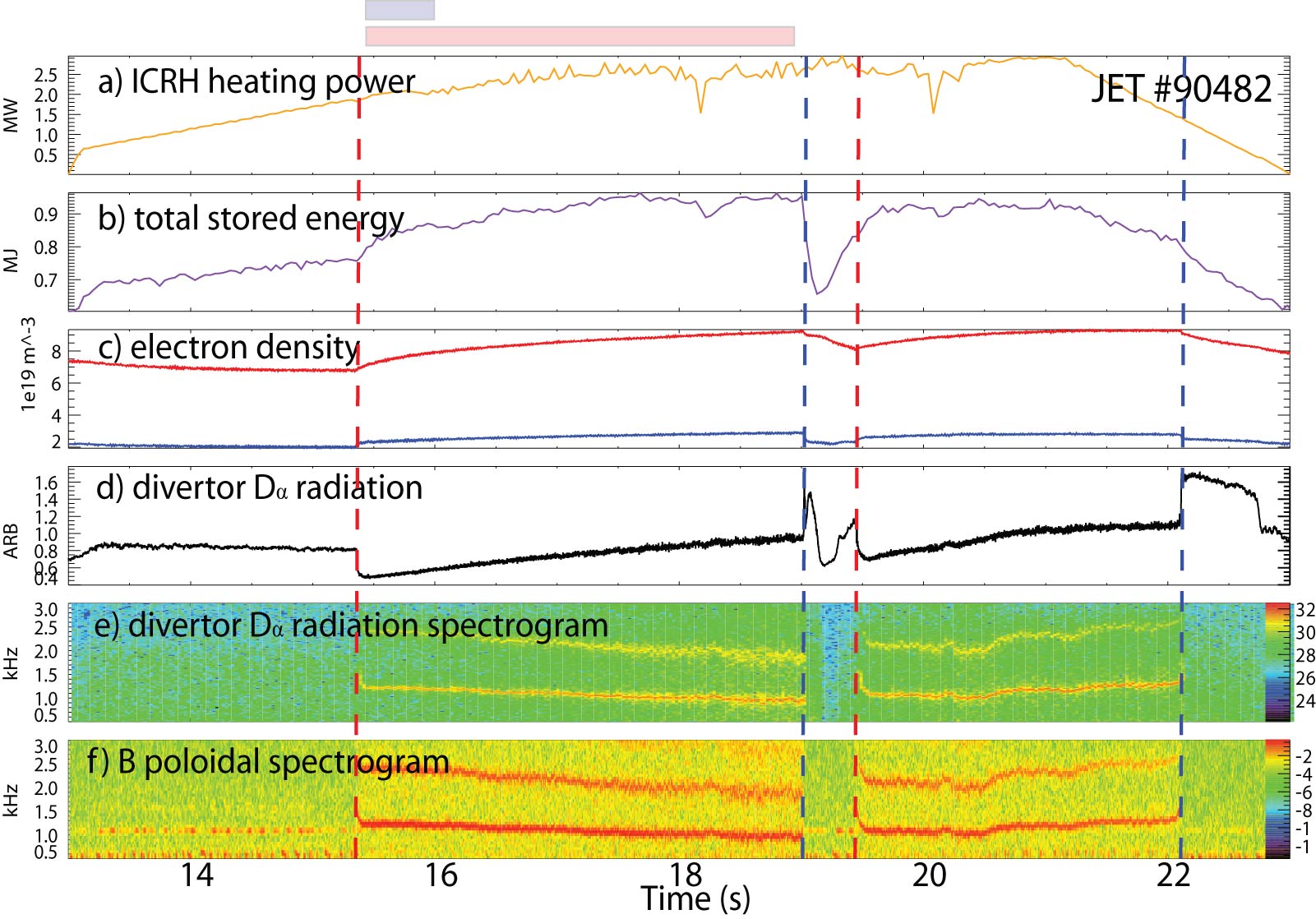} 
\caption{Overview of the analyzed JET discharge \#90482: a) ICRH heating power, b) total energy of the plasma, c) line integrated electron density from interferometer, blue - edge, red - core, d) divertor D$_\alpha$ radiation, e) spectrogram of the divertor D$_\alpha$ radiation signal, f) spectrogram of the poloidal magnetic field signal.  The L-M/M-L transition times are indicated with a vertical red/blue dashed lines. The time window for analysis shown in section~\ref{Sect.densLFO} is indicated with a blue rectangle, while for section~\ref{Sect.denshfb} is indicated with a red rectangle. \label{fig:jet.over}}
\end{figure}

\begin{figure}
\includegraphics[width=\linewidth]{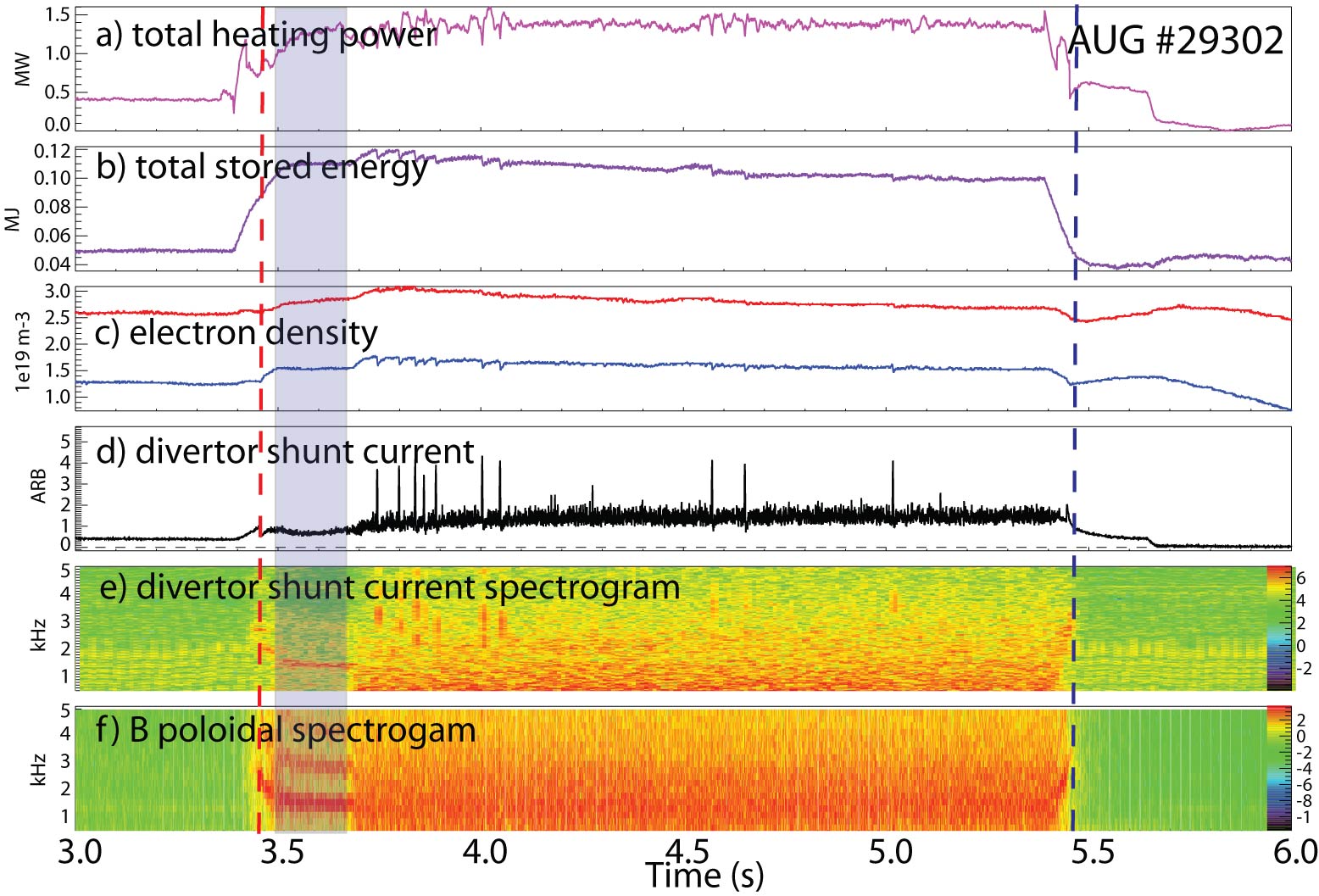} 
\caption{Overview of the analyzed AUG discharge \#29302. a) total applied power - NBI shine through - radiated power, b) total energy of the plasma, c) line integrated electron density from interferometer, blue - edge, red - core, d) divertor shunt current, e) spectrogram of the divertor shunt current signal, f) spectrogram of the poloidal magnetic field signal. The L-I/I-L transition times are indicated with a vertical red/blue dashed lines. The time window for analysis shown in section~\ref{Sect.densLFO} and in section~\ref{Sect.denshfb} is indicated with a blue rectangle. \label{fig:aug.over1}}
\end{figure}

\begin{figure}
\includegraphics[width=\linewidth]{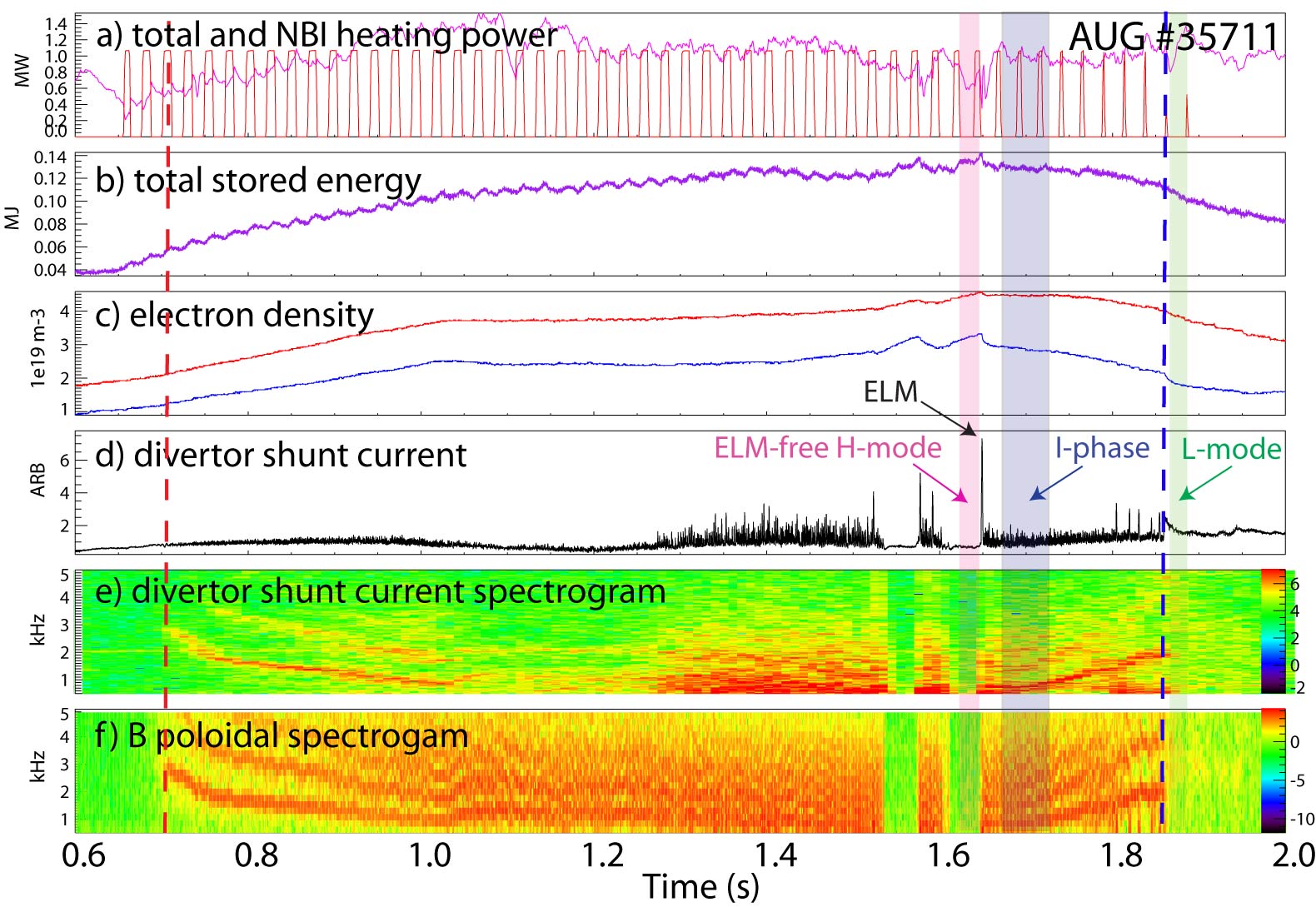} 
\caption{Overview of the analyzed AUG discharge \#35711. a) heating power (purple: total applied power - NBI shine through - radiated power) and NBI power (red) , b) total energy of the plasma, c) line integrated electron density from interferometer, blue - edge, red - core, d) divertor shunt current, e) spectrogram of the divertor shunt current signal, f) spectrogram of the poloidal magnetic field signal. The L-I/I-L transition times are indicated with a vertical red/blue dashed lines, while the time windows for ELM-free H-mode (purple), L-mode (green) and I-phase (blue) analysis time windows shown in section~\ref{Sect.presLFO} are indicated with rectangles. \label{fig:aug.over2}}
\end{figure}


The analysis presented in this paper was done for JET pulse \#90482 and \#80951, AUG pulse \#29302 and \#35711. All discharges are lower single-null configuration with favourable ion $\nabla B$ drift directed towards the x-point. The main parameters of the discharges can be seen in figure~\ref{fig:jet.over2},~\ref{fig:jet.over},~\ref{fig:aug.over1} and \ref{fig:aug.over2}. The L-M/M-L (JET) and the L-I/I-L (AUG) transition times are indicated with a vertical red/blue dashed lines in all figures.

The main parameters of JET pulse \#80951 were: $\big \langle n_e \big \rangle = 7 \times 10^{19} m^{-3}$, $B_{\phi} = 2.4 T$, $I_{p}= 2 MA$, $q_{95} \sim 3.7$. The fuelling rate was varied intentionally during the discharge for M-mode frequency scaling studies, and is reflected in the large, 1 second time scale oscillations of the total stored energy, electron density and  D$_\alpha$ radiation. The ion cyclotron resonance heating (ICRH) power had a ramp to approach the L-M transition slowly as indicated in figure~\ref{fig:jet.over2} (a). The first L-M transition is indicated with a red dashed line, and took place at about 16 s. The improved confinement is reflected in the steeper increase of the total stored plasma energy (b), increase of the plasma density both in the core and in the edge (c), and in the drop of the divertor D$_\alpha$ radiation (d). The appearance of the M-mode can be traced by the low frequency modulation of the D$_\alpha$ radiation (e), indicating the periodically changing plasma wall interaction strength. Also, the typical low frequency oscillations of the poloidal magnetic field (f), measured above the low field side (LFS) mid-plane, appeared at the transition and continues for 10 s, interrupted by an L-mode between 17.6 - 17.9 s. This discharge is a good example to show the M-mode frequency scaling ($f_{M-mode}\sim v_{Alfven,\theta}=\frac{B_{\theta}}{\sqrt{\mu_{0}m_{i}n_{e}}}$~\cite{Solano2017} ) electron density dependence due to the large edge density variation over the discharge. This discharge is used for the HFO characterization in section~\ref{Sect.HFO}.

JET pulse \#90482 was a dedicated M-mode discharge with $\big \langle n_e \big \rangle = 8.5 \times 10^{19} m^{-3}$, $B_{\phi} = 2.4 T$, $I_{p}= 2 MA$, $q_{95} \sim 3.7$. The ion cyclotron resonance heating (ICRH) power had a ramp to approach the L-M transition slowly as indicated in figure~\ref{fig:jet.over} (a). The L-M transition is indicated with a red dashed line, and took place at about 15.4 s. Subplots (b)-(e) show the same behaviour as for JET pulse \#80951.
Also, the typical low frequency oscillations of the poloidal magnetic field (f), measured below the high field side (HFS) mid-plane, appeared at the transition and continues for more than 6.5 s, interrupted by a Tungsten ablation experiment which pushes the confinement back to low confinement mode (L-mode) for about 0.5 s. M-L transitions are indicated with blue dashed lines. The analyzed time interval shown in section~\ref{Sect.densLFO} is indicated with a blue rectangle covering the 15.4 - 15.95 s time range with approximately 600 M-mode periods. The analysis shown in section~\ref{Sect.denshfb} is indicated with a red rectangle covering the 15.4 - 19.0 s time range with about 4300 M-mode periods. 

AUG pulse \#29302 (figure~\ref{fig:aug.over1}) was a dedicated fluctuation measurement programme with $\big \langle n_e \big \rangle = 3.1 \times 10^{19} m^{-3}$, $B_{\phi} = 2.5 T$, $I_{p}= 600 kA$, $q_{95} \sim 7$. The neutral beam heated (a) discharge had an L-I transition at 3.45 s, indicated by the red dashed line, I-H transition at 3.68 s and a back transition through a very short I-phase to L-mode at 5.44 s, indicated with a blue dashed line. The improved confinement is reflected in the increase of the total stored plasma energy (b), increase of the plasma density both in the core and in the edge (c), and in the drop of the divertor shunt current (d). The appearance of the I-phase can be traced by the low frequency (1.5 kHz) modulation of the divertor shunt current (e). Also, the typical low frequency oscillations of the poloidal magnetic field, measured on the HFS, below the divertor, can be followed in subplot~(f). The time interval for analysis shown in section~\ref{Sect.densLFO} and~\ref{Sect.denshfb} is indicated with a blue rectangle, covering the 3.5 - 3.7 s time range with approximately 450 I-phase periods. 

AUG pulse \#35711 (figure~\ref{fig:aug.over2}) was an L-H power threshold experiment with n = 2 magnetic perturbation (MP) at low $B_{\phi}$, with $\big \langle n_e \big \rangle = 4.6 \times 10^{19} m^{-3}$, $B_{\phi} = 1.8 T$, $I_{p}= 800 kA$, $q_{95} \sim 3.9$. A time interval prior to the application of the MP coils (1.9 s) was chosen for investigation. The I-phase appeared (indicated with a red dashed line) between 0.7 - 1.85 s when 1 MW blips of NBI heating was applied, and was interrupted twice by ELM-free H-mode parts between 1.5 - 1.7 s, both terminated by a type-I ELM, pushing the plasma back to I-phase. The back transition from I-phase to L-mode is indicated by a blue dashed line at 1.85 s, caused by the decrease of the heating power which was lowered continously from 1.5 s by shortening the NBI blips. The analyzed time intervals shown in section~\ref{Sect.presLFO} are indicated with rectangles: ELM-free H-mode (purple), L-mode (green) and I-phase (blue), the latter covering the 1.65 - 1.67 s time range with approximately 15 I-phase periods as indicated in figure~\ref{fig:cond.averaged}(d) as well.   
 
\section{Magnetic structure characterization \label{Sect.magnetic}}

\subsection{High frequency oscillation (HFO) \label{Sect.HFO}}
While the low frequency oscillations (LFO) of the magnetics were thoroughly analyzed in previous works~\cite{Solano2017}\cite{Birkenmeier2016b}, the high frequency oscillations (HFO) during the M-mode and the I-phase were not yet characterized in depth. Several bands of HFO exhibit similar temporal frequency evolution as that of the LFO. A modulation of the high frequency oscillation (10-250 kHz) power of the poloidal magnetic field is also detected during these phenomena. The HFO activity is present continuously in the magnetic signal during the L-mode, modulated during the M-mode and the I-phase, and suppressed during the developed H-mode, therefore the HFO power could be related to L-mode anomalous transport. A Type-III ELM precursor like oscillation of the radial magnetic field was detected at AUG as was shown in reference~\cite{Birkenmeier2016b}, while the poloidal field HFO power modulation was shown at JET in reference~\cite{Vianello2015}.

\begin{figure*}
\includegraphics[width=\linewidth]{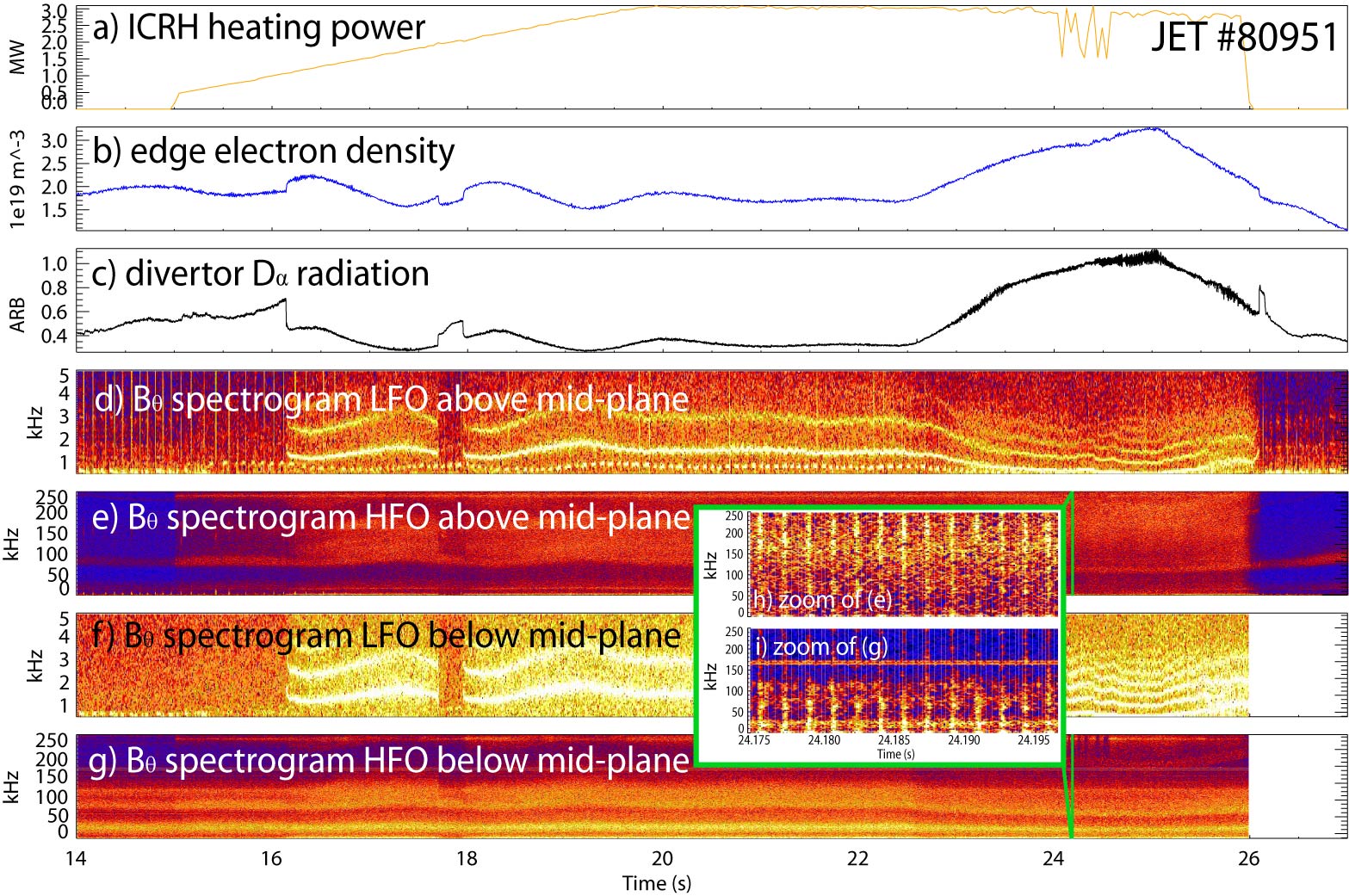} 
\caption{Heating power (a), edge electron density (b), divertor D$_\alpha$ radiation (c), spectrogram of the $\mathbf{\dot{B}_{pol}}$ signal in the low frequency (0 - 5 kHz) and the high frequency (0 - 250 kHz) range above (d, e) and below (f, g) the mid-plain, zoomed plot of the high frequency spectrograms (h, i) in the time ranges indicated with blue, for JET pulse \#80951 \label{fig:JET.HFO}}
\end{figure*}

\begin{figure*}
\includegraphics[width=\linewidth]{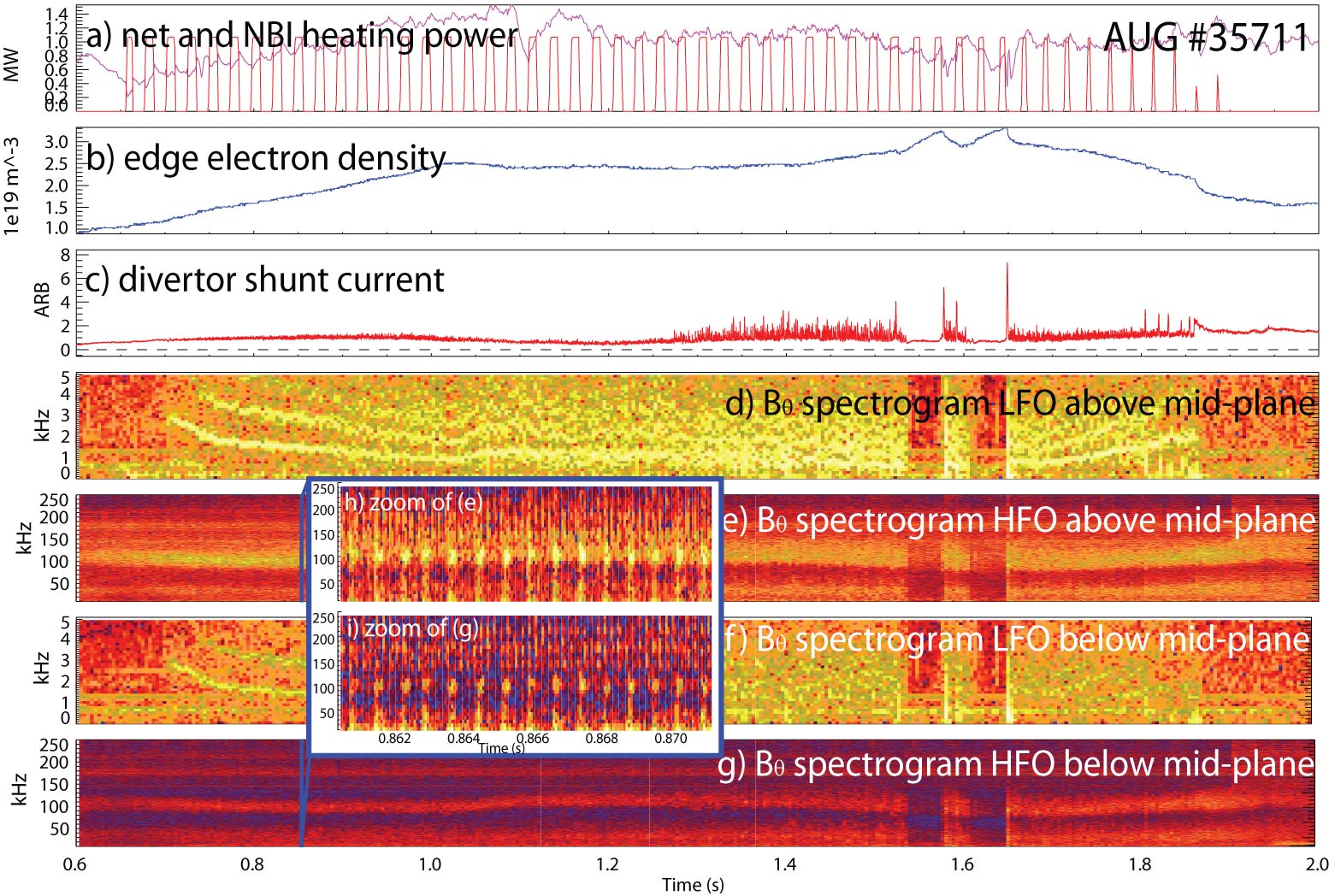} 
\caption{Heating power (a), edge electron density (b), divertor shunt current (c), spectrogram of the $\mathbf{\dot{B}_{pol}}$ signal in the low frequency (0 - 5 kHz) and the high frequency (0 - 250 kHz) range above (d, e) and below (f, g) the mid-plain, zoomed plot of the high frequency spectrograms (h, i) in the time ranges indicated with blue, for AUG pulse \#35711 \label{fig:AUG.HFO}}
\end{figure*}

The above statements about the HFO can be followed in figure~\ref{fig:JET.HFO} for the JET (pulse \#80951) and in figure~\ref{fig:AUG.HFO} for the AUG (pulse \#35711) case. The external heating power (a), the edge electron density (b), the D$_\alpha$ (JET) and the divertor shunt current (AUG) (c), the spectrogram of the $\mathbf{\dot{B}_{pol}}$ above and below the mid-plane and in the low and in the high frequency range (d, e, f, g), the zoom of the HFO spectrograms (h, i) are shown for both discharges. The time intervals of the zoomed windows are indicated by green and a blue lines for the JET and the AUG case respectively.

Several bands of HFO activity (e, g) become stronger when the external heating (a) is turned on, and weakens when it is turned off. This effect is more emphasized for the JET case. In the heated L-mode part of the discharges, most of the HFO bands are present with stationary frequency. The L-M/M-L and the L-I/I-L transitions can be identified by the drop/rise of the D$_\alpha$ (\ref{fig:JET.HFO}c), by the steeper increase/decrease of the edge density (b) signals, and also by the appearance of the LFO (d, f). The divertor shunt current signal (\ref{fig:AUG.HFO}c) does not show a steep drop at the L-I transition due to the slowly ramped heating, however the I-L transition is clearly visible after a few dithering transitions at 1.85 s indicated by the approximately doubled signal amplitude. The frequency of the HFO bands (e, g) follow the changes of the LFO (d, f) frequency when the plasma is in the M-mode / I-phase. The modulation of the HFO activity can be followed in the subplots (h, i) which appears simultaneously in all coil signals.

\subsection{High frequency oscillation power modulation \label{Sect.HFB}}

\begin{figure*}
\includegraphics[width=\linewidth]{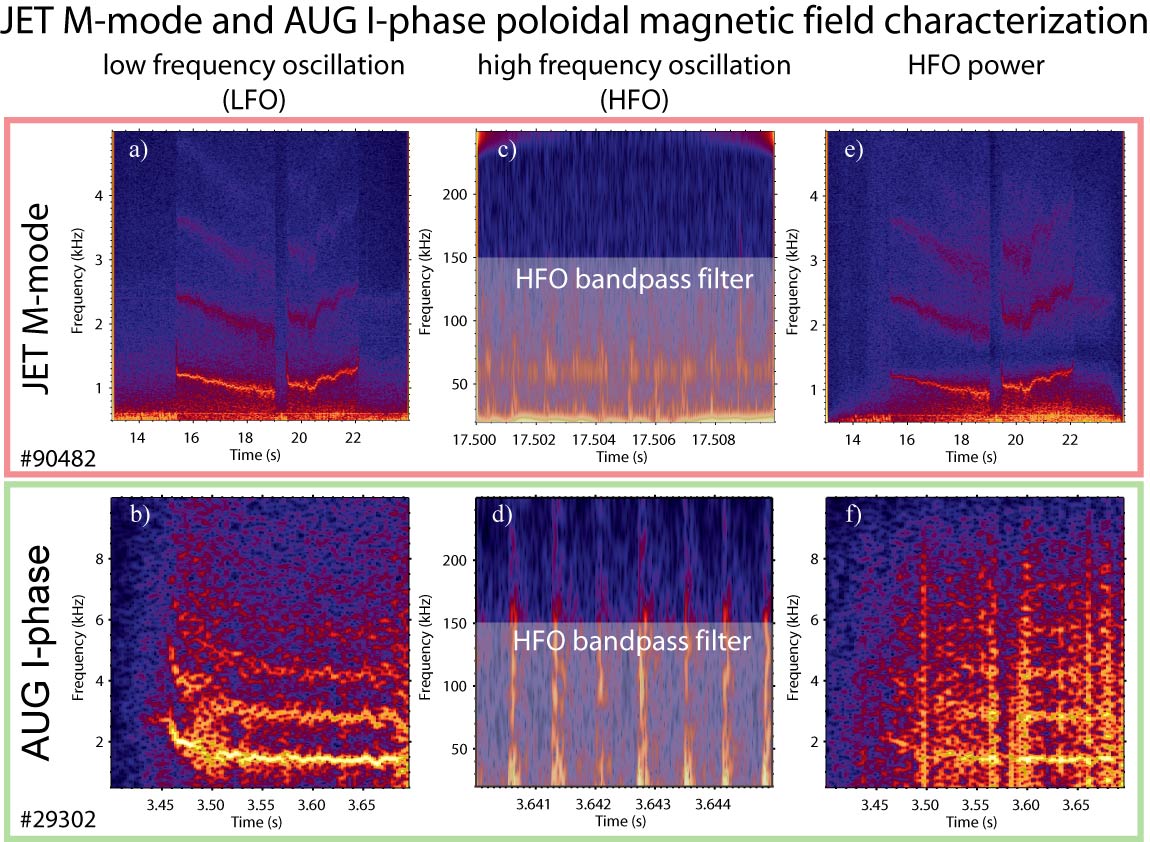} 
\caption{Low frequency range spectrogram of $B_{pol}$ for M-mode (a) and I-phase (b), high frequency range spectrogram of $B_{pol}$ for M-mode (c) and I-phase (d) and $B_{pol}$ HFO power modulation for M-mode (e) and I-phase (f). The frequency range for HFO power calculation is indicated with a white rectangle in (c) and (d). The time trace of the HFO power modulation is identical that of the low frequency oscillation (LFO). (JET: discharge \#90482, Mirnov coil below HFS mid-plane; AUG: discharge \#29302, Mirnov coil below HFS divertor) \label{fig:bpol.fluct}}
\end{figure*}
The modulation of the HFO activity of the magnetic signals raises the question, if this modulation is present continuously during the M-mode and the I-phase. An analysis is carried out for both the JET (discharge \#90482) and the AUG (discharge \#29302) cases to show that in both machines the LFO and the HFO power modulation are correlated. The plots were produced with the FLIPP (Wigner RCP) and the NTI Wavelet Tools~\cite{Pokol2019}\cite{Pokol2013} (Budapest University of Technology) codes. The HFO power modulation signal is calculated by taking the square of the poloidal magnetic field signal after filtering with a 10-150 kHz band-pass digital finite impulse response (FIR) filter (indicated by the white rectangles in Figure~\ref{fig:bpol.fluct} (c) and (d)), which gives the time evolution of the HFO power in the signal. (A signal after similar processing can be seen in figure~\ref{fig:cond.averaged}(d) for AUG discharge \#35711.) Figure~\ref{fig:bpol.fluct} shows the spectrogram of the magnetic signal in the low frequency (a, b) and the high frequency (c, d) range, and the spectrogram of the HFO power modulation (e, f) for the M-mode and the I-phase cases respectively. The time evolution of the LFO and the HFO power modulation frequency is identical, the signals also show high coherence in the low frequency range. This indicates that the HFO power modulation in the magnetic signals is present during the whole M-mode and the whole I-phase, and are closely coupled with the LFO. The relative phase of the HFO power modulation was found to be 0 between several coil pairs and numerous discharges for both the M-mode and the I-phase. This indicates that the HFO power modulation is happening simultaneously in all coil signals where it can be detected. These properties are observed on both machines, and these HFO power signals are used as reference signal in the further analysis.

\subsection{Toroidal symmetry \label{Sect.modenumber}}

\begin{figure*}
\includegraphics[width=\linewidth]{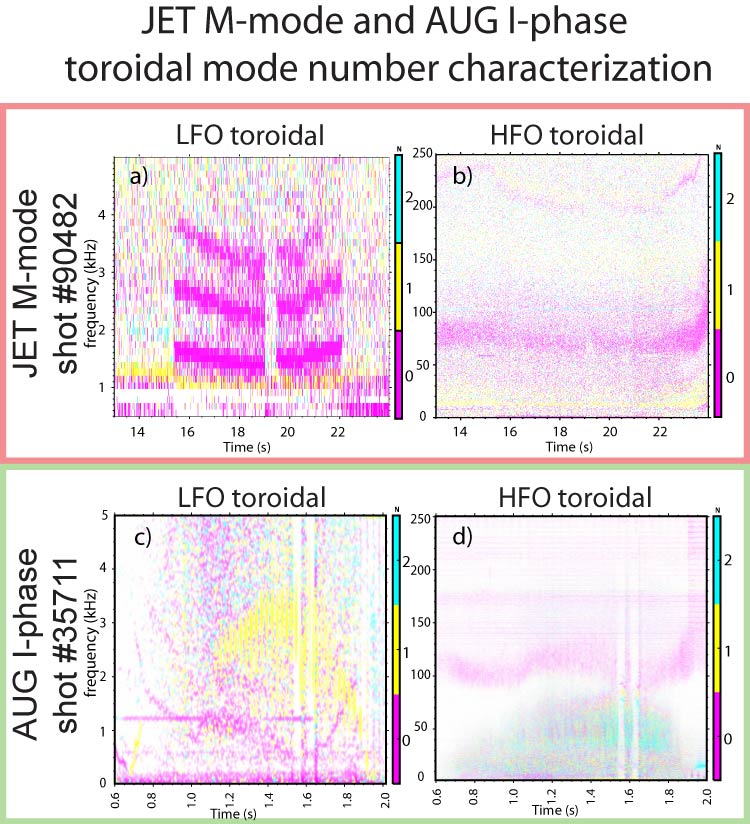} 
\caption{Spectrogram with the dominant toroidal mode numbers indicated in the low frequency (a) and the high frequency (b) range (JET: discharge \#90482, all available coils for toroidal mode number estimation); Spectrogram with the dominant toroidal mode numbers indicated in the low frequency (c) and the high frequency (d) range (AUG: discharge \#35711, all available coils were used) \label{fig:modenumber}}
\end{figure*}

The toroidal symmetry estimation of the HFO was carried out for both machines. (The HFO poloidal symmetry estimation was attempted and turned out to be out of the scope of this paper.) Figure~\ref{fig:modenumber} summarizes the results of the analysis: upper row for JET while the lower the AUG. The mode number calculation is based on a fitting method in the toroidal and poloidal position - relative phase plane for each time-frequency domain. The color depth in figure~\ref{fig:modenumber} corresponds to the coherence level, while the color to the best mode number fit for each domain in the spectrograms. The plots were produced with the PySpecview (AUG) and the Spec Mode (JET) codes. 

The toroidal mode number can be easily identified both in the low (a, c) and the high frequency (b, d) range for the frequency bands which follow the M-mode/I-phase frequency evolution. The results confirm the n=0 symmetry of the LFO, and suggests n=0 symmetry of the HFO both for the M-mode and the I-phase in the band close to 100 kHz. 

Table~\ref{table:magnetic} summarizes the findings about the magnetic symmetries of the M-mode and the I-phase: the LFO is m=1, n=0; the HFO is n=0; HFO power modulation happens simultaneously in every coil signal where it can be detected.

\begin{table}[h]
\caption{\label{table:magnetic}Magnetic symmetry properties of the M-mode and the I-phase.}\begin{indented}
\item[]\begin{tabular}{@{}cccc@{}}
\br
& \textbf{LFO} & \textbf{HFO} & \textbf{\begin{tabular}[c]{@{}c@{}}HFO \\ power modulation\end{tabular}} \\ 
\mr
\textbf{JET M-mode}  & m=1, n=0     & n=0          & simultaneousy in every signal \\
\textbf{AUG I-phase} & m=1, n=0     & n=0          & simultaneousy in every signal \\ 
\br
\end{tabular}
\end{indented}

\end{table}

\section{Density profile oscillations \label{Sect.densLFO}}

\begin{figure*}
\includegraphics[width=\linewidth]{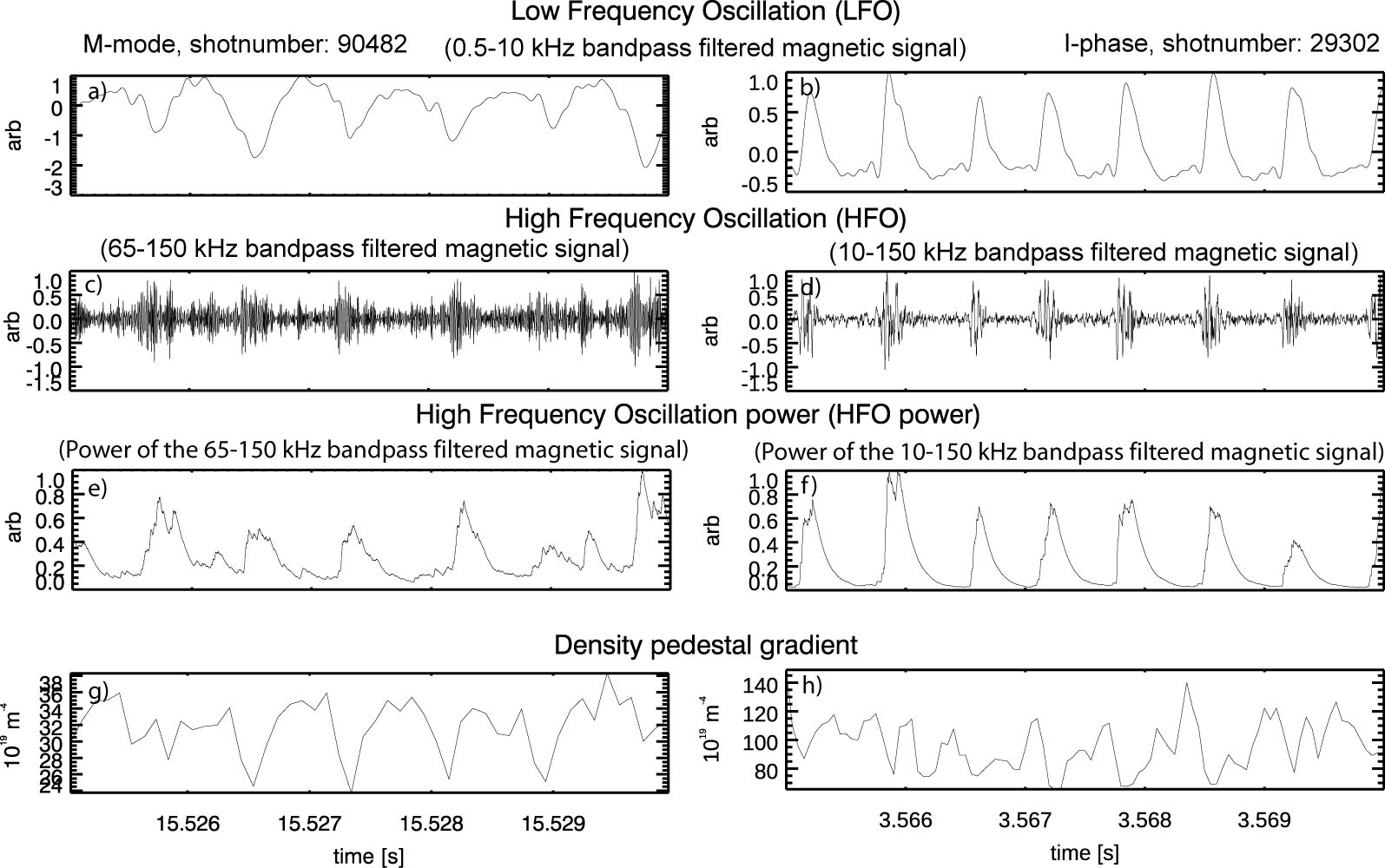} 
\caption{Time traces of the LFO - low frequency bandpass filtered magnetic signal for M-mode (a) and I-phase (b); HFO - high frequency bandpass filtered magnetic signal for M-mode (c) and I-phase (d); HFO power - power of the high frequency bandpass filtered magnetic signal for M-mode (e) and I-phase (f) and the density pedestal gradient for M-mode (g) and I-phase (h) \label{fig:pedgrad_combined}}
\end{figure*}

The density profile is calculated from Li-BES light emission profile data using a Bayesian algorithm \cite{Fischer2008} for JET (discharge \#90482) and for AUG (discharge \#29302). Modulation related to the studied phenomena is analyzed two ways to have a clear interpretation. The first approach can be seen in figure~\ref{fig:pedgrad_combined} showing the time traces of a low frequency bandpass filtered (0.5 - 10 kHz) magnetic signal (LFO) in subfigure (a) and (b), the high frequency bandpass filtered (65 - 150 kHz and 10 - 150 kHz respectively) magnetic signal (HFO) in (c) and (d), the power of the high frequency bandpass filtered signal (HFO power) in (e) and (f) and the density pedestal gradient in (g) and (h) for the M-mode and for the I-phase respectively. The density gradient was evaluated as follows: the each density profile was fitted with a tangent hyperbolic function and the pedestal gradient was calculated as the pedestal height over the pedestal width. The pedestal gradient shows same periodicity as the other signals, the pedestal collapses after the HFO activity starts, but a statistical approach is necessary since the uncertainty of the measurement is comparable with the fluctuation amplitude caused by the phenomenon.

\begin{figure*}
\includegraphics[width=\linewidth]{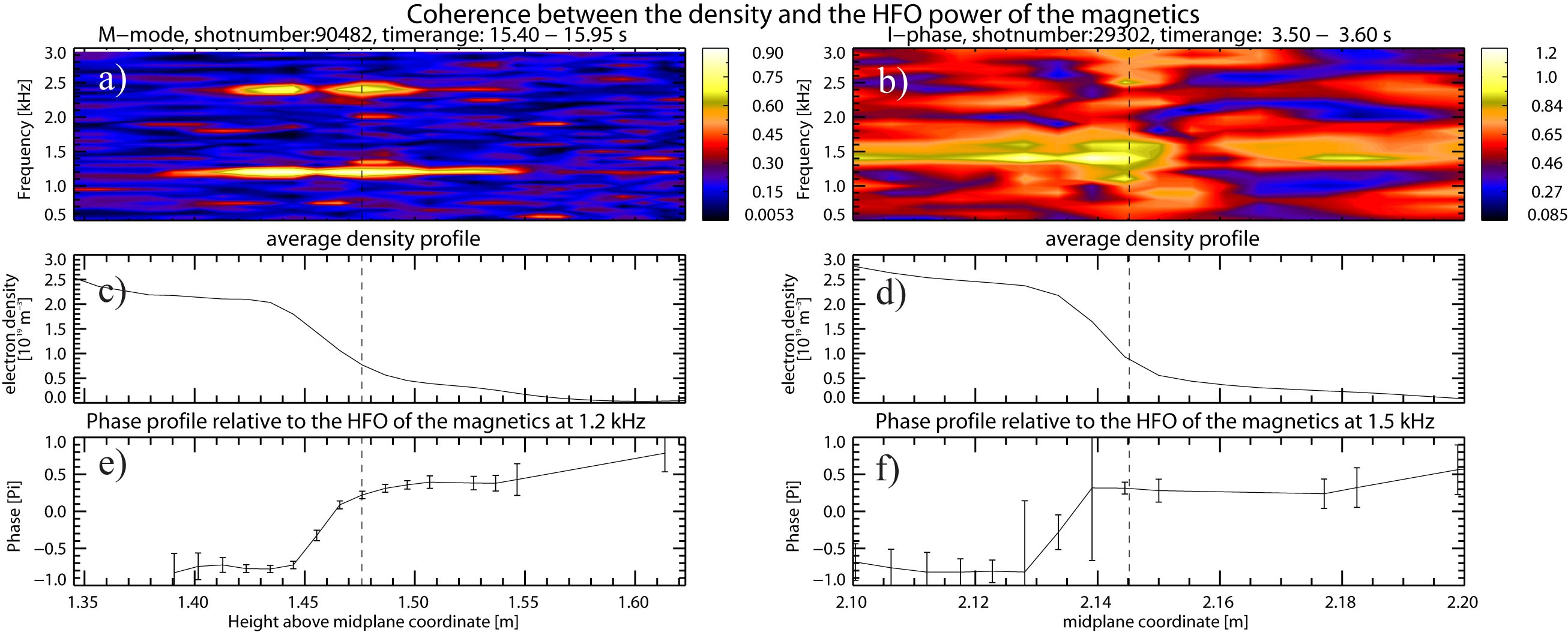} 
\caption{M-mode (a) and I-phase (b) coherence of the HFO power modulation of the magnetics and the LFO of the electron density; M-mode (c) and I-phase (d) averaged density profile; M-mode (e) and I-phase (f) phase profile at the relevant frequency range relative to the HFO power modulation of the magnetics, points with uncertainty higher than $\pi \over 2$ were omitted. Vertical dashed line indicates the LCFS position. \label{fig:dens.coherence}}
\end{figure*}

To get a precise picture of the temporal ordering between the profile dynamics and the turbulence a statistical analysis has been carried out. The M-mode and the I-phase related density profile modulation is analyzed in terms of coherence spectra of a magnetic HFO power modulation signal as reference and the time evolution of the reconstructed density at different radial locations along the Li-BES beam path. Figure~\ref{fig:dens.coherence} shows the general density profile behavior of the M-mode on the left, I-phase on the right: the (a) and (b) figures show the coherence, the (c) and the (d) the average density profile while (e) and (f) the phase profile relative to the HFO power modulation. The coordinates are not translated to normalized poloidal flux coordinates to avoid the possible inaccuracies from the magnetic reconstruction, rather the physical coordinates are used: height above mid-plane coordinates for the JET, and mid-plane coordinates for AUG matching the beam injection directions. The top and the bottom of the pedestal is mostly modulated, as indicated by the two maxima in the coherence at these positions at the M-mode/I-phase frequency. The relative phase between the top and bottom of the pedestal fluctuation is $\pi$, while the middle of the pedestal is less affected, which indicates that the gradient is modulated, as was shown by the ultra-fast reflectometer as well in reference~\cite{Medvedeva2017}. The relative fluctuation amplitude is about 10\% at the pedestal top and 20\% at the pedestal bottom in both cases (not shown in these figures, but has been calculated). The phase of the pedestal bottom density relative to the HFO power modulation of the magnetics is +0.3$\pi$ which means that the flattening of the pedestal is preceded by the HFO pulse in the magnetics by 120 $\mu$s. The negative (-0.7$\pi$) phase at the pedestal top represents the same: the minima of the pedestal top density during the flattening of the pedestal is preceded by the HFO activity maxima by 0.3$\pi$. A radially outward propagating density perturbation in the SOL is also related to the phenomena, indicated by the clear phase delay outwards in the SOL. Note, that these results are fairly similar for the M-mode and the I-phase.

\section{Temperature and density profile characterization at AUG \label{Sect.presLFO}}

\begin{figure*}
\includegraphics[width=\linewidth]{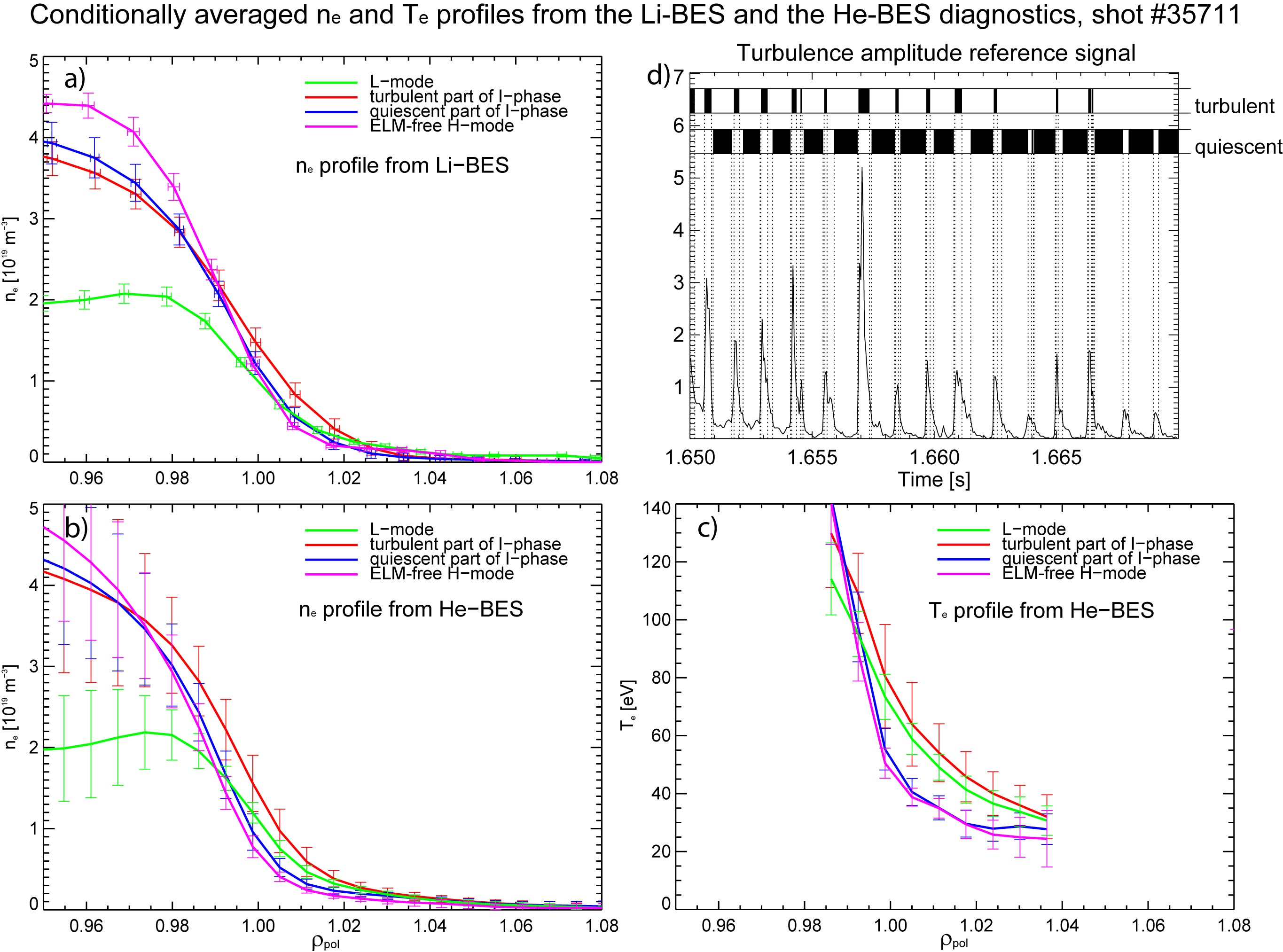} 
\caption{Conditionally averaged Li-BES density profile (a), He-BES density profile (b) and He-BES temperature profile (c) as a function of $\rho_{pol}$ for AUG discharge \#35711. Each plot contains four conditionally averaged profiles: L-mode (green), turbulent I-phase (red), quiescent I-phase (blue) and ELM-free H-mode (purple), the error-bars show the standard deviation of the data and the position values. The reference signal for the conditional averaging (d) is the high frequency band (50 - 250 kHz) power of a poloidal field coil signal (see section~\ref{Sect.magnetic}) which can be considered as the turbulence amplitude. The time intervals for the conditional averaging is shown above with `turbulent' and `quiescent' labels. \label{fig:cond.averaged}}
\end{figure*}

Several plasma edge instabilities are excited above a critical pressure gradient, thus the question if the temperature profile is modulated and if in phase with the density raises naturally. The He-BES diagnostic at AUG measures the SOL and plasma edge temperature and density profile simultaneously (see section~\ref{Sect.He-BES}). The analysis was carried out at AUG (discharge \#35711, see section~\ref{Sect.dataset}). 

(The analysis has been carried out on the electron cyclotron emission data at AUGs, but the edge channels were found to be affected by shine-through according to electron cyclotron forward modeling~\cite{Vanovac2018}\cite{Denk2018}, thus no solid conclusions could have been drawn from that diagnostic.)

Conditionally averaged density profile from Li-BES (a), density profile from He-BES (b) and temperature profile from He-BES (c) as a function of $\rho_{pol}$ are shown in figure~\ref{fig:cond.averaged}. In this paper, $\rho_{pol}$ refers to the normalized poloidal magnetic flux coordinate. Each plot contains four profiles according to the time intervals in figure~\ref{fig:aug.over2}: L-mode (green), turbulent I-phase (red), quiescent I-phase (blue) and ELM-free H-mode (purple), the error-bars show the standard deviation of the data and the position values. The reference signal for the conditional averaging (d) is the high frequency band (50 - 250 kHz) power of a poloidal field coil signal, as calculated in section~\ref{Sect.magnetic}, which can be considered as the turbulence amplitude. The time intervals for the conditional averaging is shown above the signal with `turbulent' and `quiescent' labels. The He-BES $T_{e}$ data is only plotted in a restricted radial range: the $T_{e}$ reconstruction out of a singlet and a triplet line fails in the $\rho_{pol} > 1.03$ region due to the applied static collisional radiative model~\cite{Griener2018}, which does not properly handle the dynamic population process of the triplet spin system of helium in low $n_{e}$ regions. This however does not affect the $n_{e}$ evaluation, as here only singlet lines are used. For $\rho_{pol} < 0.98$ the signal to noise ratio drops due to beam attenuation.

The $n_{e}$ profile gradient in the plasma edge region ($0.98 < \rho_{pol} < 1.0$) during the turbulent part of the I-phase is comparable to that of the L-mode, while during the quiescent part of the I-phase it is close to that of the ELM-free H-mode according to the Li-BES data. The steep density pedestal region is wider during the H-mode which results in higher pedestal top density compared to the quiescent part of the I-phase. The He-BES diagnostic located at a different toroidal and poloidal location, however indicates similar $n_{e}$ profile behavior which suggests a m=0, n=0 density modulation of the I-phase. The decreasing values of the $n_{e}$ and $T_{e}$ in the near SOL region ($1.0 < \rho_{pol} < 1.3$) indicate increasing confinement as investigating L-mode - turbulent I-phase - quiescent I-phase - H-mode profiles in sequence. 
The $T_{e}$ values in the near SOL during the quiescent I-phase and H-mode are significantly lower than during the turbulent I-phase and L-mode, indicating the H-mode confinement properties of the quiescent I-phase part. 
Finding a lower SOL temperature in H-mode than in L-mode needs some explanation, since the separatrix temperature in steady-state H-mode is higher compared to L-mode, considering H-modes are usually heated more than L-modes. The ELM-free H-mode part we used is highly transient, and the heating power did not change considerably through the transitions, thus the power deposited in the SOL will be defined by the energy loss from the confined region. This indeed reduced in ELM-free H-mode due to the transport barrier formation, resulting in a colder separatrix relative to the L-mode case. Also, it can be stated based on the He-BES data, that the $n_{e}$ and the $T_{e}$ profile is modulated in phase, that is the I-phase modulates the electron pressure gradient. It has to be noted, that the observations about the $n_{e}$ dynamics is very similar to that of the COMPASS LCO phenomenon as was shown in reference~\cite{Grover2018}.

\section{Temperature profile oscillations at JET \label{Sect.tempLFO}}

\begin{figure}
\includegraphics[width=\linewidth]{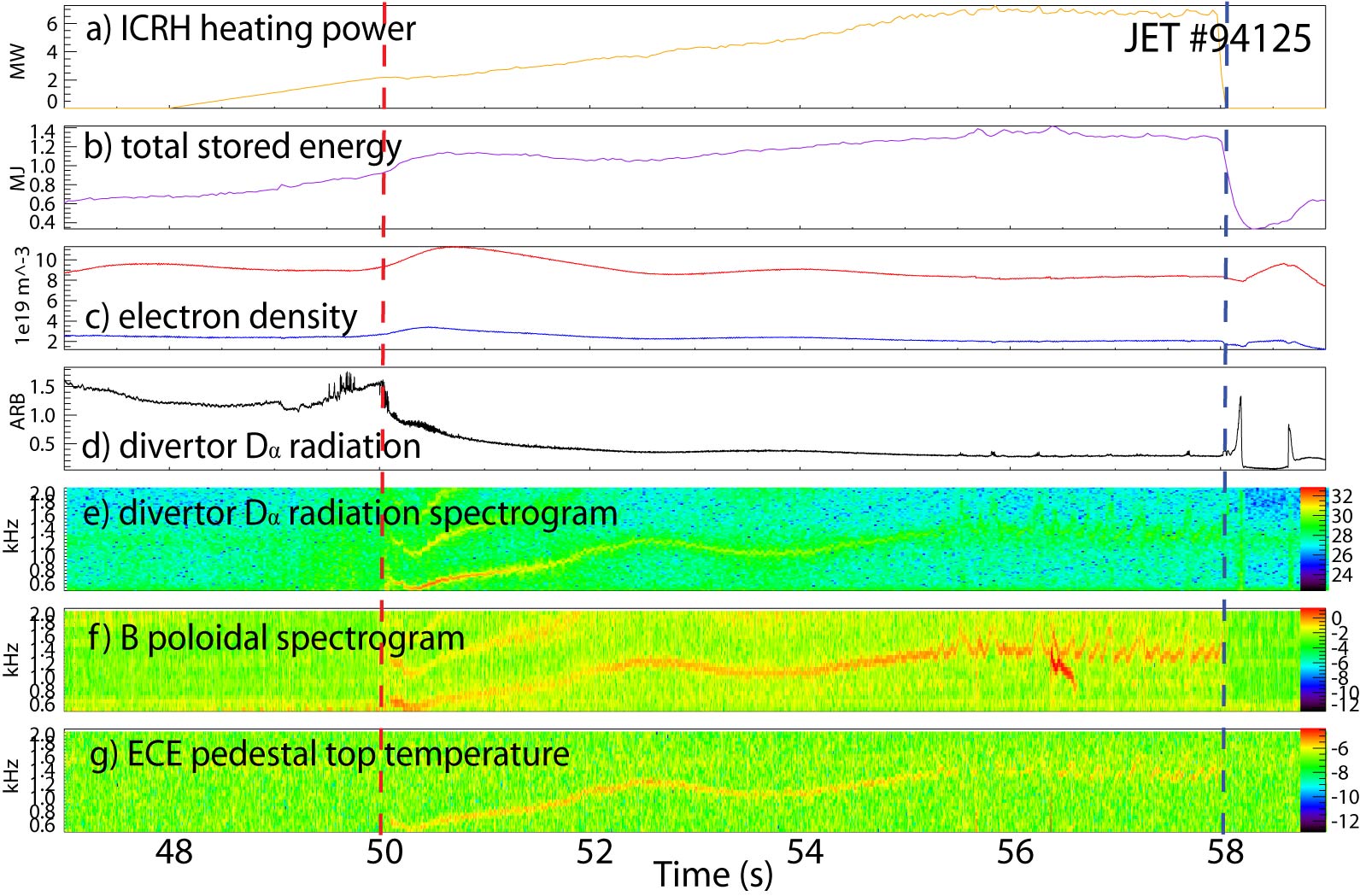} 
\caption{Overview of the analyzed JET discharge \#94125: a) ICRH heating power, b) total energy of the plasma, c) line integrated electron density from interferometer, blue - edge, red - core, d) divertor D$_\alpha$ radiation, e) spectrogram of the divertor D$_\alpha$ radiation signal, f) spectrogram of the poloidal magnetic field signal, g) spectrogram of the pedestal top temperature signal from the ECE diagnostic.  The L-M/M-L transition times are indicated with a vertical red/blue dashed lines.  \label{fig:jet.over3}}
\end{figure}

The temperature profile fluctuations were investigated at JET analyzing the electron cyclotron emission (ECE) radiometer data for discharge \#94125. The pedestal top channels of the ECE data exhibit identical fluctuations as the divertor D$_\alpha$ and the poloidal magnetic field signal as can be seen in figure~\ref{fig:jet.over3} (e), (f) and (g). The pedestal top density was sufficiently high ($3 - 4 \times 10^{19} m^{-3}$) to exclude shine-through effects. The pedestal top electron temperature is modulated by the JET M-mode, however no conclusions can be drawn about the phase relative to the density, due to the low spatial resolution of the ECE and the uncertainty of the EFIT mapping.

\section{Density HFO power modulation \label{Sect.denshfb}}

\begin{figure*}
\includegraphics[width=\linewidth]{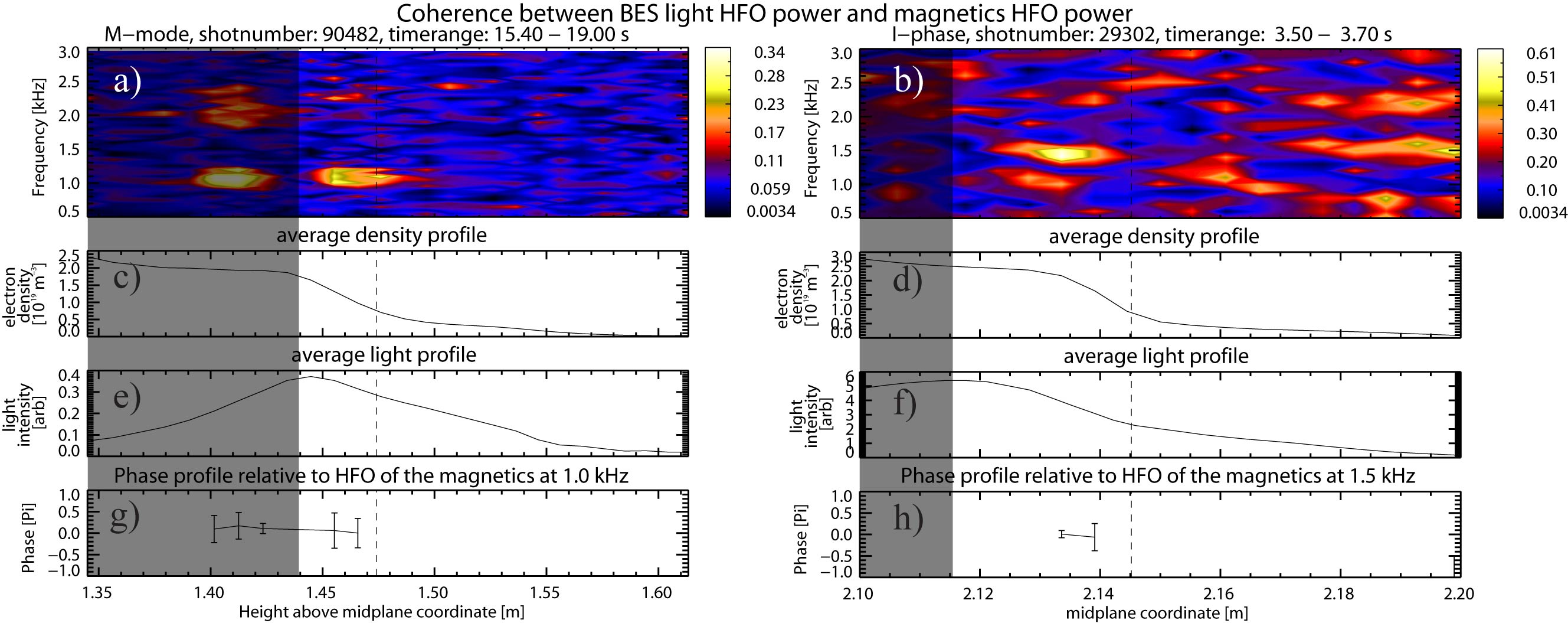} 
\caption{M-mode (a) and I-phase (b) coherence of the HFO power modulation of the magnetics and the HFO power modulation of the Li-BES light emission; M-mode (c) and I-phase (d) averaged density profile; M-mode (e) and I-phase (f) averaged light profile; M-mode (g) and I-phase (h) phase profile at the relevant frequency range relative to the HFO power modulation of the magnetics, points with uncertainty higher than $\pi \over 2$ were omitted. Vertical dashed line indicates the LCFS position. The gray rectangle shows the region where the localization of the fluctuation response is affected by the atomic physics of the beam. \label{fig:light.hfb.coherence}}
\end{figure*}

The HFO power modulation of the magnetics raises the question if there is any detectable high frequency density modulation related to the M-mode and the I-phase. The below analysis show that the HFO is present in the density at the pedestal region.

The low signal to noise ratio (5 - 10) of JET and AUG Li-BES measurements does not allow us to directly detect the density fluctuations in the high frequency range since the spectra of the beam emission signals are dominated by noise above 50 kHz. However, the characteristic frequency of the HFO power modulation (see section~\ref{Sect.magnetic}) enables us to localize radially the HFO activity using correlation techniques. The HFO (10-50 kHz) power modulation calculation was carried out on the Li-BES light emission signals. The time range for the analysis was 15.4 - 19 s in the JET (discharge \#90482) and 3.5 - 3.7 s in the AUG (discharge \#29302) case as shown in figure~\ref{fig:jet.over}.

Figure~\ref{fig:light.hfb.coherence} summarizes the results for the M-mode on the left, I-phase on the right: (a) and (b) show the coherence with the HFO power signal, (c) and (d) the average density profile, (e) and (f) the average light emission profile while (g) and (h) the phase profile relative to the HFO power signal of the magnetics in the frequency range of interest. Phase points with uncertainty higher than $\pi \over 2$ were omitted. The coherence plots show significant peaks at the M-mode and the I-phase frequency, considering the 0.09 and the 0.18 significance levels respectively. This means that the HFO power is locally modulated with the characteristic mode frequency. \ref{App.fluct1} gives deeper insights about the Li-BES fluctuation measurement interpretation while \ref{App.fluct2} about the Li-BES fluctuation response modeling efforts to support the statements of this section.


The AUG case is clear where the high coherence peak at the I-phase frequency is on the increasing edge of the light profile and located at the pedestal, between 2.13 - 2.14 m. In the JET case, the second peak (counting from the beam injection location which is on the right) between 1.39 - 1.42 m is on the falling edge of the light emission profile and is due to the beam attenuation effect as described in the appendix. The blind spot is between 1.43 - 1.45 m close to the light emission profile maxima. The first peak between 1.45 - 1.49 m is on the rising edge of the light profile and located at the pedestal region. Only this first peak corresponds to real density fluctuation, the second is due to beam effects (see \ref{App.fluct1} and \ref{App.fluct2}). The HFO power modulation of the Li-BES signal and of the electron density accordingly is localized in the pedestal region in both the JET and the AUG cases. The phase relative to the HFO power modulation of the magnetic signal is close to 0 implying that the bursts in the magnetics and in the density at the pedestal are simultaneous and are probably related to each other. Note, that these results are fairly similar for the M-mode and the I-phase.

\section{Summary and discussion\label{Sect.summary}}

In summary we conclude, that the M-mode and the I-phase have very similar properties in terms of density profile dynamics and magnetics signatures. The following observations during both phenomena have been made, based on the analysis of the Mirnov coil, the Li-BES and the ECE signals:

\begin{enumerate}
\item The low frequency oscillation (LFO) of the magnetic signals has m=1, n=0 symmetry. This was shown previously for the M-mode~\cite{Solano2017} and for the I-phase~\cite{Birkenmeier2016b}.
\item A high frequency oscillation (HFO) of the magnetic signals is seen in broad frequency bands all the time during the L-mode, it is suppressed periodically during the M-mode and the I-phase and suppressed continously during the H-mode thus the HFO power could be related to L-mode anomalous transport.
\item The dominant frequency bands of the HFO of the magnetic signals show similar temporal evolution as the LFO frequency. The frequency scaling of the M-mode and the I-phase  (see some discussion below) was out of the scope of this paper, thus it is a qualitative statement about the observation, that when the LFO frequency increases/decreases the dominant HFO frequency increases/decreases as well. 
\item The HFO of the magnetic signals has n=0 symmetry in the band close to 100 kHz. 
\item The HFO power of the magnetic signals is modulated with the LFO frequency, and the modulation appears simultaneously in all coil signals. The latter four points confirm prior speculation shown for the M-mode~\cite{Vianello2015}.
\item The HFO power of the $n_{e}$ is modulated with the LFO frequency at the pedestal region. 
\item The HFO power modulation of the $n_{e}$ happens simultaneously with the HFO power modulation of the magnetic signals. The latter two point are entirely new results.
\item The $n_{e}$ profile flattens after the burst of the HFO in the density and the magnetics. This was shown for AUG in for example~\cite{Cavedon2017} but the detailed profile dynamics is a new result for both JET and AUG.
\item The $T_{e}$ is modulated with the LFO frequency at the pedestal region. 
\item The flattening of the $n_{e}$ profile is followed by an outward propagating density pulse in the SOL. The latter two points confirm prior results shown in \cite{Solano2017} for the M-mode and in \cite{Birkenmeier2016b},\cite{Cavedon2017} for the I-phase.
\end{enumerate}

Additionally, a combined analysis of the He-BES and the Li-BES data was carried out during the I-phase at the AUG, investigating the $T_{e}$ and the $n_{e}$ profiles. This analysis confirmes previous results shown in for example~\cite{Cavedon2017}. The following observations were made:

\begin{enumerate}[resume]
\item Both the $T_{e}$ and the $n_{e}$ profiles flatten after the burst in the HFO power of the magnetics and the density.
\item The $T_{e}$ and the $n_{e}$ profiles are modulated in phase in the near SOL, indicating that the electron pressure ($p_{e}$) gradient is modulated by the I-phase.
\item The $n_{e}$ profile is modulated in phase at different poloidal and toroidal locations suggesting toroidal and poloidal symmetry of the density modulation.
\end{enumerate}

The growth of the turbulence amplitude in the pedestal region along with HFO activity in the magnetics refers to MHD mode de-stabilization, which is followed by the degradation of the confinement and the flattening of the pressure profile. The HFO activity stops due to the lack of drive likely from the free-energy in the profile gradients, the pressure pedestal builds back up, and the cycle starts over. This observation leads to the conjecture that the I-phase and the M-mode are not governed by predator-prey dynamics, since the confinement improvement through zonal flow generation would be expected to happen after the phase with high turbulence activity. This is clearly not the case here. This is in understanding with the prior work documented on AUG in reference~\cite{Cavedon2017} and is not in contradiction with the results in reference~\cite{Medvedeva2017} since our results are based on averaged data in stationary I-phase.

Another puzzling question to be clarified in upcoming works is the difference in the frequency scaling of the M-mode and I-phase, in particular \cite{Solano2017} claims that the M-mode frequency does not scale with the ion sound speed while \cite{Birkenmeier2016b} finds that the frequency of the I-phase depends on the temperature. As to our understanding, the differences in frequency scaling can be related to the speed of pedestal formation, which could be quite different in the fast transient I-phase evaluated at AUG and in the almost steady M-modes evaluated at JET.

It is also clear, that both phenomena exhibit electromagnetic nature since the LFO, the HFO and the HFO power modulation of the magnetics and the density are coupled. Our analysis revealed that the M-mode phenomena at JET and the I-phase phenomena at AUG have very similar dynamics, thus their physical background is the same.

\ack

This work has been carried out within the framework of the EUROfusion Consortium and has received funding from the Euratom research and training programme 2014-2018 and 2019-2020 under grant agreement No 633053. The views and opinions expressed herein do not necessarily reflect those of the European Commission.

 \section*{References}

 \bibliographystyle{iopart-num}
 \bibliography{M-mode}
 
\appendix
\section{Li-BES fluctuation measurement interpretation \label{App.fluct1}}
The interpretation of the results of section~\ref{Sect.denshfb} necessitates the review of the effects of the applied statistical methods and the Li-BES techniques. The noise modulation by light intensity modulation, the light emission response to density fluctuations and the spatial localization of the measurement are to be discussed. 

\begin{itemize}
\item
The low frequency light modulation causes light HFO power modulation as well, since the noise is proportional to the square root of the signal amplitude in case of quantities with Poisson statistics. This case, namely that the coherence peaks are caused by noise amplitude modulation can be excluded, since the light LFO and the light HFO power modulation are not in phase.
\item 
The Li-BES light emission fluctuations are usually considered proportional to the local density fluctuations, however, the fluctuation response is gradually decreasing inwards the plasma~\cite{Willensdorfer2014}. Reaching a point where the beam attenuation and excitation terms cancel each other, the beam emission has no response to density fluctuations, and is called the blind spot accordingly. Over that point the light response is negative to local density increase. Moreover, at any point along the beam path the beam carries information about all previous processes, e.g. density fluctuation at the beginning of the beam evolution causes negative response over this point through ionization losses. The gray rectangles indicate regions in figure~\ref{fig:light.hfb.coherence} where the results are not reliable due to this effect.
\item 
The Li-BES systems measure the line emission of Lithium atoms that were excited by the plasma particles. The spontaneous decay of the valence electrons has a finite half life of approximately 27 ns which causes 2 cm spatial smearing of the light emission since the beam atoms are traveling in a 50 keV mono-energetic beam. If the collisional excitation and de-excitation processes from the observed line transition are taken into account, which are proportional to the local electron density, the smearing decreases as the beam penetrates into the higher density region. Therefore, the fluctuation response will be more localized at higher density as shown in reference~\cite{Asztalos2017}. 
\end{itemize}

\section{Li-BES fluctuation response modeling \label{App.fluct2}}

\begin{figure}
\centering
\includegraphics[width=10cm]{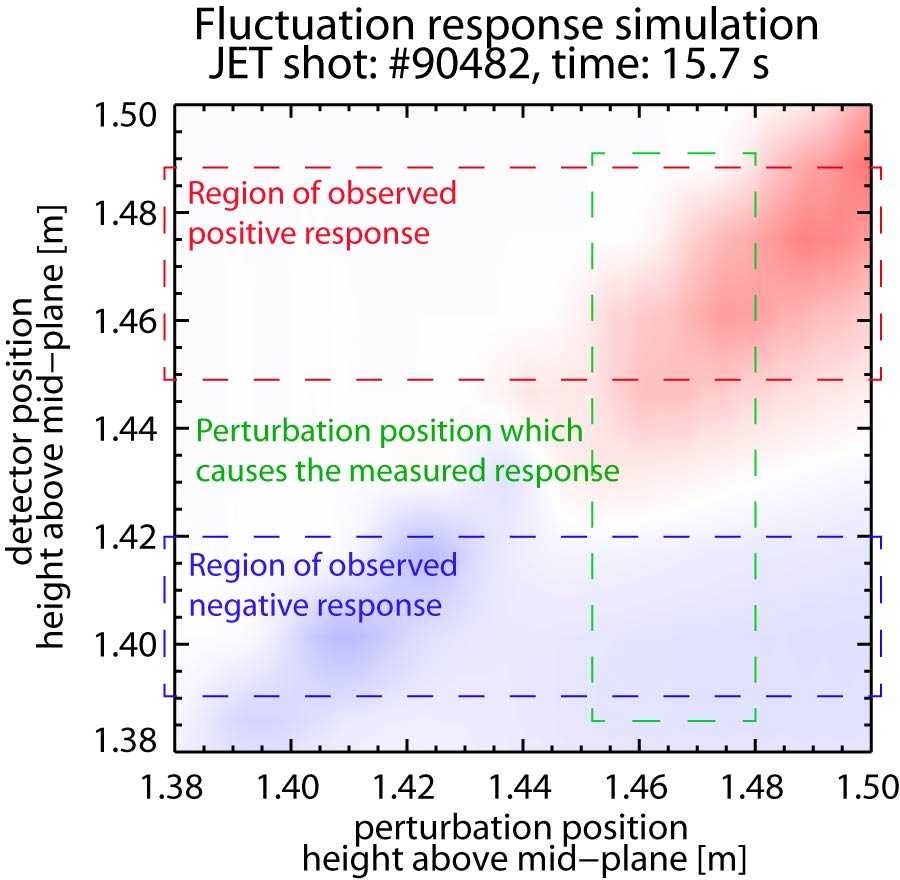} 
\caption{Fluctuation response simulation: the contour plot shows the light response at a position indicated on the y axis for a density perturbation at a position indicated on the x axis both in height above mid-plane coordinate (same as for the JET case in figure~\ref{fig:light.hfb.coherence}). The regions of the observed positive and negative responses are indicated with red and blue dashed rectangles respectively. The region where a density perturbation causes the measured light fluctuation response is highlighted with a green rectangle. \label{fig:fluctresponse.renate}}
\end{figure}

A fluctuation response simulation has been carried out with the RENATE BES simulator~\cite{Guszejnov2012} to support the statement about the density perturbation localization for the JET case. The simulation places three dimensional density perturbations along the beam line step-wise, aligned with the magnetic field lines. Beam evolution calculation is performed on the updated density profile and the light profile is acquired. The detected photon current on each detector pixel and for each perturbation is calculated, accounting for the three dimensional observation system modeling \cite{Asztalos2017}. By subtracting the equilibrium light profile from the perturbed light profiles, the calculation gives the light response on each detector for a density perturbation at different locations. The result can be seen in figure~\ref{fig:fluctresponse.renate}, and reads as follows. The contour plot shows the light response at a position indicated on the y axis for a density perturbation at a position indicated on the x axis. The positive response is red, the negative is blue. In our case, as can be seen in figure~\ref{fig:light.hfb.coherence}(a), a positive response is located in the 1.45 - 1.49 m range (indicated with a red dashed rectangle here), while a negative response is located in the 1.39 - 1.42 m range (indicated with a blue dashed rectangle here). The location of the perturbation causing such response can be read from the plot, indicated with a green dashed rectangle, and falls in the pedestal region.
The measured phase is 0 at both locations in figure~\ref{fig:light.hfb.coherence} which seems to be in contradiction with the expected $\pi$ phase jump due to the positive response at one while negative response at the other location implied by the modeling. In reality the HFO is in counter phase at the 2 locations, however we are investigating its power modulation which erases the phase information.

\end{document}